\documentclass[sigconf, nonacm]{acmart}

\AtBeginDocument{%
  }

\usepackage[utf8]{inputenc} 
\usepackage[T1]{fontenc}    
\usepackage{hyperref}       
\usepackage{url}            
\usepackage{booktabs}       
\usepackage{amsfonts}       
\usepackage{enumitem}       
\usepackage{nicefrac}       
\usepackage{microtype}      
\usepackage{xcolor}         
\usepackage{graphicx} 
\usepackage{multirow}
\usepackage{hyperref}
\usepackage{amsmath}
\usepackage{geometry}
\usepackage{xcolor} 
\usepackage{colortbl} 
\usepackage{booktabs} 
\usepackage{todonotes}

\usepackage{dcolumn}
\usepackage{bm}
\usepackage{algorithm2e}

\usepackage{braket}

\usepackage{subcaption}
\usepackage{tikz}
\usetikzlibrary{quantikz2}
\usepackage{tabularx}
\usepackage{multirow}
\usepackage{array}
\usepackage{balance}

\newcommand{\red}[1]{{\color{red}\texttt{#1}}}
\newcommand{\blue}[1]{{\color{blue}\texttt{#1}}}

\begin{document}

\title{Medusa: Detecting and Removing Failures for Scalable Quantum Computing}

\author{Karoliina Oksanen}
\email{karoliina.i.oksanen@aalto.fi}
\affiliation{%
  \institution{Aalto University}
  \city{Espoo}
  \country{Finland}
}

\author{Quan Hoang}
\email{quan.hoang@aalto.fi}
\affiliation{%
  \institution{Aalto University}
  \city{Espoo}
  \country{Finland}
}

\author{Alexandru Paler}
\email{alexandru.paler@aalto.fi}
\affiliation{%
  \institution{Aalto University}
  \city{Espoo}
  \country{Finland}
}

\renewcommand{\shortauthors}{Oksanen et al.}

\begin{abstract}
Quantum circuits will experience failures that lead to computational errors. We introduce Medusa, an automated compilation method for lowering a circuit's failure rate. Medusa uses flags to predict the absence of high-weight errors. Our method can numerically upper bound the failure rate of a circuit in the presence of flags, and fine tune the fault-tolerance of the flags in order to reach this bound. We assume the flags can have an increased fault-tolerance as a result of applying surface QECs to the gates interacting with them. We use circuit level depolarizing noise to evaluate the effectiveness of these flags in revealing the absence of the high-weight stabilizers. Medusa reduces the cost of quantum-error-correction (QEC) because the underlying circuit has a lower failure rate. We benchmark our approach using structured quantum circuits representative of ripple-carry adders. In particular, our flag scheme demonstrates that for adder-like circuits, the failure rate of large-scale implementations can be lowered to fit the failure rates of smaller-scale circuits. We show numerically that a slight improvement in the local fault-tolerance of the flag-qubits can lead to a reduction in the overall failure rate of the entire quantum circuit. 
\end{abstract}




\maketitle

\section{Introduction}

Quantum error-correction is usually applied to computations without taking into account the structure of the corresponding circuit. This approach, although straightforward, leads to high overheads in terms of qubit counts and execution time. It is not obvious how to account for the structure of the circuit while compiling it to fault-tolerant, error-corrected primitives. It is known that arbitrary computations can be rewritten as measurement-based ones~\cite{briegel2009measurement, vijayan2024compilation,bombin2024unifying}, where a cluster state (i.e. stabilizers state which is classically efficient to simulate) is consumed through measurements.

The advantage of flags is that they have low-overhead compared to alternative QEC methods when it comes to circuit depth and qubit count. One specific use case is to use flags to detect errors on stabilizer measurements with stabilizer codes.~\cite{chao2020flag}  In this way the flags detect dangerous errors i.e. those that would propagate to higher weights in the circuit. 

The concept of flags has been further used to implement e.g. fault-tolerant state preparation for both quantum error correction codes~\cite{peham2025automated} and magic states~\cite{chamberland2019fault}. For stabilizer codes the practical implementation of flag schemes and challenges related to them have also been explored, e.g. limitations set by qubit connectivity.~\cite{cho2024accelerated, zen2024quantum}. Additionally optimization of flag schemes when it comes to circuit depth and parallelization has been studied~\cite{liou2025reducing, bhatnagar2023low}.

The failure rate of a circuit, often called logical error rate, can be defined in many ways. In~\cite{delfosse2025low} stabilizers were used to lower the failure rate of a quantum circuit, where the failure rate represents the probability that errors will propagate in such a manner that the output of the circuit has a non-trivial error. To improve the logical failure rate of a circuit, based on stabilizer measurements, a circuit may be restarted. 

\begin{figure}[!t]
    \begin{subfigure}{0.45\columnwidth}
        \centering
        \begin{tikzpicture}
            \node[scale=0.7]{
                \begin{quantikz}[row sep={0.6cm,between origins}, column sep=0.34cm]
                \lstick{$\ket{\psi}$} & \targ{} & \meter{} \\
                \lstick[2]{$\ket{+}$} & \ctrl{-1} && \meter{} \\
                & & \gate{X}\wire[u][2]{c} & \gate{Z}\wire[u][1]{c} & \rstick{$\ket{\psi}$}
                \end{quantikz}
            };
        \end{tikzpicture}
        \caption{ }
        \label{fig:ICM}
    \end{subfigure}
    \begin{subfigure}{0.45\columnwidth}
        \centering
        \begin{tikzpicture}
        \node[scale=0.7]{
            \begin{quantikz}[row sep={0.4cm,between origins}, column sep=0.34cm]
            \lstick{$\ket{+}$} & \ctrl[style={draw=blue, fill=blue}]{3} & \ctrl{2}\gategroup[3,steps=3,style={dashed,rounded corners,fill=gray!20, inner xsep=2pt},background]{} & \push{\red{X}} & \ctrl{1} & \ctrl[style={draw=blue, fill=blue}]{3} & \rstick{\red{X}} \\
            \lstick{$\ket{0}$} &&&& \targ{} && \rstick{\red{X}} \\
            \lstick{$\ket{+}$} && \targ{} &&&& \\
            \lstick{$f_x, \ket{0}$} & \targ[style={draw=blue}]{} &&&& \targ[style={draw=blue}]{} & \rstick{\red{X}}
            \end{quantikz}
        };
        \end{tikzpicture}
        \caption{}
        \label{fig:exb}
    \end{subfigure}
    \caption{Quantum circuits: (a) \textsc{Medusa} is based on ICM circuits~\cite{paler2017fault, vijayan2024compilation}, e.g. gate teleportation, which consist entirely of CNOT gates, initializations and measurements; (b) A bit flip error occurring in location \textit{X} in the original circuit is caught by the flag: the measurement of the flag would give 1 instead of 0 as the error is propagated to the flag through the blue CNOT gate. At the end of the circuit the error is present on the first, second and last qubit. By post-selecting based on the flag measurements and checking the stabilizers of the original ICM circuit we can examine wether or not the flags have successfully caught the errors.}
    \label{fig:example}
\end{figure}
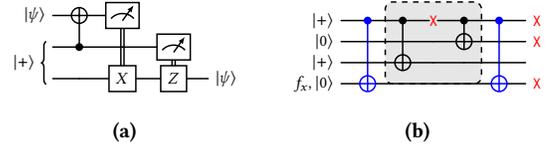

\subsection{Background}

\emph{Flags} are qubits that are added to a pre-existing quantum circuit in order to detect errors. Using two CNOT operations between a flag qubit and a pre-existing qubit of the flagless quantum circuit, errors occurring between the two CNOTS can be detected with a final readout measurement of the flag qubit. This measurement is then used to perform post selection as described in Sec.~\ref{sec:ler}.

\begin{figure}[!h]
    \begin{subfigure}{0.49\columnwidth}
        \begin{tikzpicture}
        \node[scale=0.7]{
            \begin{quantikz}[row sep={0.4cm,between origins}, column sep=0.34cm]
            \lstick{$q_0$} & \ctrl[style={draw=blue, fill=blue}]{3} & \ctrl{1} & \ctrl{2} &  \ctrl[style={draw=blue, fill=blue}]{3} &\\
            \lstick{$q_1$} && \targ{} &&&\\
            \lstick{$q_2$} &&& \targ{} &&\\
            \lstick{$f_x$} & \targ[style={draw=blue}]{} &&& \targ[style={draw=blue}]{} & \meter{}
            \end{quantikz}
        };
        \end{tikzpicture}
        \caption{}
        \label{fig:exa}
    \end{subfigure}
    \begin{subfigure}{0.49\columnwidth}
        \begin{tikzpicture}
        \node[scale=0.7]{
            \begin{quantikz}[row sep={0.4cm,between origins}, column sep=0.34cm]
            \lstick{$q_0$} && \targ[style={draw=blue}]{} & \targ{} & \targ{} & \targ[style={draw=blue}]{} &&\\
            \lstick{$q_1$} &&& \ctrl{-1} &&&&\\
            \lstick{$q_2$} &&&& \ctrl{-2} &&&\\
            \lstick{$f_z$} & \gate[style={draw=blue}]{\color{blue}H} & \ctrl[style={draw=blue, fill=blue}]{-3} &&& \ctrl[style={draw=blue, fill=blue}]{-3} & \gate[style={draw=blue}]{\color{blue}H} & \meter{} 
            \end{quantikz}
        };
        \end{tikzpicture}
        \caption{}
        \label{fig:exz}
    \end{subfigure}
    \caption{Flag circuits obtained after adding a flag qubit and gates, marked blue: (a) an X-flag;  (b) a Z-flag.}
    \label{fig:flagExample}
\end{figure}
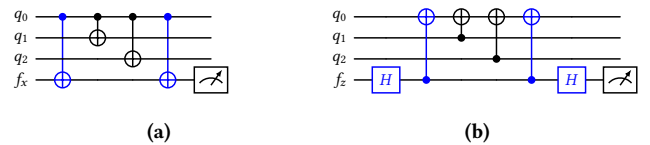

The flags are separated into two categories, X- and Z-flags, which are used to detect bit and phase flip errors respectively (Fig.~\ref{fig:flagExample}) Flags are added by attaching a new ancilla qubit (i.e. flag qubit) to a quantum circuit, and a readout measurement of 1 indicates that errors might have occured within the circuit.

With flags, the failure rate can be defined either with or without post-selection. If no post-selection is used, the flag outcomes are used to correct the errors which occur~\cite{chao2018fault}. However, one can still use the flag outcomes to improve the error rate without error correction by using post-selection~\cite{debroy2020extended}. This means that the circuits are not restarted or corrected in the presence of errors but rather some results are discarded completely. In this case the goal is to predict wether or not errors are present based on the flags.

\subsection{Problem Statement: Constrained Compilation of Flag Circuits}

Circuit compilation is usually performed under a set of constraints, such as limited number of qubits or gates. Assuming the limited availability of resources and an upper bound for error rates, how can we make the best use of flags? Alternatively, for achieving a given target failure rate, how should we catch errors of the highest weight possible using a limited number of flags? Answering this question is a an optimization problem which we decompose into two sub-problems: in Sec.~\ref{sec:add} we focus on \emph{can simple flag adding heuristics be efficient in lowering a circuit's failure rate?}, and in Sec.~\ref{sec:tune} we continue with \emph{is it possible to increase the fault-tolerance of the flags and how is this reflected in the resources needed?}.

Synthesizing optimal quantum circuits given a set of constraints is very challenging, especially for large circuits. Nevertheless, there is a rich literature on applying SAT solvers for the optimal compilation of quantum circuits. For example, the method of~\cite{schneider2023sat} is a combination of an SAT solver and binary search for finding optimal circuits. The optimization was done with respect to two-qubit gate count. The approach is based on representing quantum circuits with a combination of tableaus~\cite{aaronson2004improved}, storing information on which gates are present in the circuit~\cite{berent2022towards}. In other words, quantum states are represented by the operations that generate them rather than a set of amplitudes. In~\cite{peham2025automated} an SAT approach was used to synthesize Clifford circuits for the preparation of fault-tolerant states.

SAT solvers have not been shown to compile circuits of more than seven qubits~\cite{shaik2025cnot} and heuristic approaches have been presented as an alternative~\cite{peham2025automated}. For example, there are $\mathcal{O}(n^2/\log n)$ gates needed to optimally implement any $n$-qubit Clifford circuit~\cite{aaronson2004improved}. Assuming the generators of the Clifford gate set ($H, S$ and $CNOT$ gates), the total number of binary variables needed to represent a Clifford circuit synthesis problem is $\mathcal{O}(n^5/\log n)$~\cite{schneider2023sat}.


\begin{figure}
    \includegraphics[width=\columnwidth]{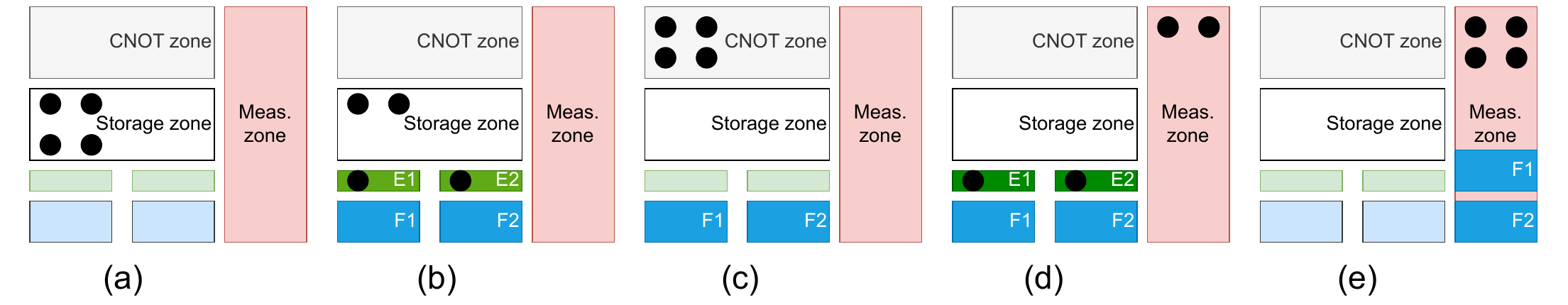}
    \caption{\textsc{Medusa} for zoned architectures. Illustration for a hypothetical 4 qubit circuit (black circles): (a) The qubits are stored in the storage zone; (b) in order to add two flags to the circuit, two of the qubits are moved to a green encoding zone and are encoded into cat states (dark green), and the resulting encoded state is performing the first CNOT with a fault-tolerant flag (dark blue rectangle) residing in flag zone (previously light blue); (c) after unencoding the physical qubits, these are moved to a CNOT zone where they are executing the ICM circuit; (d) the flagged qubits are moved back to the entangling zones to perform the second CNOT necessary for connecting the flag; (e) finally, the qubits and the flags are moved to the measurement zone.}
    \label{fig:arch}
\end{figure}

\subsection{Contribution}

There is interest in automating the process of inserting flags~\cite{zen2024quantum, acharya2021automated}, but the automatic placing of flags into arbitrary circuits has not yet been widely studied. This work introduces \textsc{Medusa}, an automated method for reducing failure rates in large quantum circuits using uniquely assigned flags. Our main contributions are: 1) \textbf{Failure rate framework.} We formalize failure, failure rate, and post-selected failure rate using stabilizer comparisons between noisy and noiseless ICM circuits. 2) \textbf{Scalable flag insertion.} We propose a structure-based heuristic that identifies and ranks and inserts unique flag candidates by weight; 3) \textbf{Upper-bound estimation.} By simulating circuits with perfect flags, we determine numerical upper bounds on achievable failure rates improvements for each circuit family. 4) \textbf{Flag fault-tolerance tuning.} We introduce an error multiplier $m$ and use a binary-search heuristic to find the minimal flag reliability needed to reach a target FR (e.g., matching the FR of an $(N\!-\!1)$-sized circuit); 5) \textbf{Resource implications.} Assuming flags are protected by surface codes, we map required $m$ values to code distances and qubit counts, yielding practical hardware estimates.

Additionally, \textsc{Medusa} can be easily adapted and used on quantum computers supporting zoned architectures~\cite{ransford2025helios, henriet2020quantum,bluvstein2025fault} (Fig.~\ref{fig:arch}). Due to the high cost of full error-correction, it is possible by adding fault-tolerant (partially error-corrected) flags to lower the total failure rate of a computation and to enable early fault-tolerant quantum computations~\cite{eisert2025mind,preskill2025beyond}.

\begin{table*}[t!]
\renewcommand{\arraystretch}{.9}
\centering
\small
\setlength{\tabcolsep}{2pt}  

\begin{tabular}{
| >{\centering\arraybackslash}m{0.06\linewidth} 
|| >{\centering\arraybackslash}m{0.17\linewidth} 
| >{\centering\arraybackslash}m{0.27\linewidth} 
|| >{\centering\arraybackslash}m{0.17\linewidth} 
| >{\centering\arraybackslash}m{0.27\linewidth} |} 
\hline
\multirow{2}{3.5em}{\centering\arraybackslash\textbf{Input}} & \multicolumn{2}{c ||}{\textbf{Fault-less circuit} (Fig.~\ref{fig:exa})} & \multicolumn{2}{c |}{\textbf{Faulty circuit} (Fig.~\ref{fig:exb})} \\ 
\cline{2-5}
  & \textbf{Stabilizers} & \textbf{Stabilizers + Flags} & \textbf{Stabilizers} & \textbf{Stabilizers + Flags} \\ 
\hline
\texttt{000} & \texttt{+ZII, +IZI, +IIZ} & \texttt{+ZIII, +IZII, +IIZI, +IIIZ} & \texttt{-ZII, -IZI, +IIZ} & \texttt{-ZIII, -IZII, +IIZI, -III\red{Z}} \\ 
\texttt{+00} & \texttt{+XXX, +ZIZ, +IZZ} & \texttt{+XXXI, +ZIZI, +IZZI, +IIIZ} & \texttt{+XXX, -ZIZ, -IZZ} & \texttt{+XXXI, -ZIZI, -IZZI, -III\red{Z}} \\ 
\texttt{++0} & \texttt{+XIX, +ZIZ, +IXI} & \texttt{+XIXI, +ZIZI, +IXII, +IIIZ} & \texttt{+XIX, -ZIZ, +IXI} & \texttt{+XIXI, -ZIZI, +IXII, -III\red{Z}} \\ 
\texttt{+0+} & \blue{\texttt{+XXI, +ZZI, +IIX}} & \texttt{\blue{+XXII, +ZZII, +IIXI,} +IIIZ} & \blue{\texttt{+XXI, +ZZI, +IIX}} & \texttt{\blue{+XXII, +ZZII, +IIXI,} -III\red{Z}} \\ 
\texttt{0+0} & \texttt{+ZII, +IXI, +IIZ} & \texttt{+ZIII, +IXII, +IIZI, +IIIZ} & \texttt{-ZII, +IXI, +IIZ} & \texttt{-ZIII, +IXII, +IIZI, -III\red{Z}} \\ 
\texttt{0++} & \texttt{+ZII, +IXI, +IIX} & \texttt{+ZIII, +IXII, +IIXI, +IIIZ} & \texttt{-ZII, +IXI, +IIX} & \texttt{-ZIII, +IXII, +IIXI, -III\red{Z}} \\ 
\texttt{00+} & \texttt{+ZII, +IZI, +IIX} & \texttt{+ZIII, +IZII, +IIXI, +IIIZ} & \texttt{-ZII, -IZI, +IIX} & \texttt{-ZIII, -IZII, +IIXI, -III\red{Z}} \\ 
\texttt{+++} & \blue{\texttt{+XII, +IXI, +IIX}} & \texttt{\blue{+XIII, +IXII, +IIXI,} +IIIZ} & \blue{\texttt{+XII, +IXI, +IIX}} & \texttt{\blue{+XIII, +IXII, +IIXI,} -III\red{Z}} \\ 
\hline
\end{tabular}
\vspace{12pt}
\caption{The stabilizers which are checked for the circuit in Fig.~\ref{fig:exa} with different input states, with and without flags. Blue stabilizers do not change when an error is present. Red measurements catch errors on the flags.}
\label{table:stabilizers}
\end{table*}

\section{Methods}

To compile circuits larger than the ones presented in~\cite{schneider2023sat, peham2025automated} (e.g. $>100$ physical qubits), while avoiding the scalability challenges of SAT-based methods, we propose a heuristic that takes a circuit's structure into account to add flags (Sec.~\ref{sec:add}). Afterwards, we improve the circuit's failure rate by tuning the flags (Sec.~\ref{sec:tune}).

The effectiveness of the flags can be quantified by analyzing the circuit's failure rates, and in order to do this we start from the stabilizers of the circuit. By post-selecting based on the flag measurements and checking stabilizers of the original circuit, we examine whether or not the flags will successfully catch errors.

First, without loss of generality, we consider arbitrary computations pre-compiled to ICM~\cite{paler2017fault, vijayan2024compilation}. Such circuits, which are related to graph states, consist entirely of CNOT gates (Sec.~\ref{sec:circuits}). This enables the tracking, the simulation and the analysis of failures and their rates. In order to evaluate the failure rate of an ICM circuit, we start from a truth-table like representation of the circuit's stabilizers (Table~\ref{table:stabilizers}). Moreover, ICM circuits allow the straightforward insertion of flags with the sole purpose of detecting failures~\cite{chao2018quantum, chao2018fault}.

Second, starting from the stabilizer table, we informally define the failure rate as the result of asking: \emph{How often does the result of the measurements differ from the expected ones, when the circuit is under noise?}. To this end, we distinguish between flagless ICM circuits (i.e. containing no flags) and flagged ICM circuits (i.e. containing flags introduced automatically), and would like to use flags for determining when a set of stabilizer measurements is different from the expected ones.

Third, we simulate each flagless ICM circuit in the absence of noise. At the end of the circuit, a set of canonical stabilizers is computed using Stim~\cite{gidney2021stim} and we use these stabilizers to build the truth table.

Fourth, we numerically upper bound the failure rate achievable with the introduced flags. We perform noisy simulations of both the flagless and flagged ICM circuit and measure the stabilizers. We determine that an error has occurred, if the stabilizer measurement of the noisy flagged circuit are different from the stabilizers of the noiseless flagless circuit. It has been noted that measuring the full set of stabilizers may be expensive as the size of the set can be quite large~\cite{delfosse2025low, peham2025automated}. For the circuit sizes we worked with this was still plausible but for larger circuits following the approach in~\cite{delfosse2025low} and selecting \textit{r} random stabilizers to check instead may be a better option.

Finally, we adapt the fault-tolerance of the flags to be as close as possible to the numerically established upper bound.

\subsection{Failure Rate}
\label{sec:ler}

Herein, we formalise the definitions of the various failure rates discussed throughout the manuscript. All the definitions assume the existence of a stabilizer truth table similar to the one from Table~\ref{table:stabilizers}.

\textbf{Definition - Failure:} A failure of a circuit occurs when measuring the canonical stabilizers of a circuit returns a result different from $\ket{0}$. 

\textbf{Definition - Failure Rate (FR):} The failure rate of a flag-less circuit is given by the number of times where the stabilizer measurements differed from the expected ones, divided by the total number of measurements. 

In the case of a flag circuit, the failure rate can be defined with or without post-selection: with post-selection only those cases are chosen for analysis where no error has been detected based on the results of flag measurement. This means that some cases are completely discarded and thus not included in calculating the failure rate. Without post-selection, the failure rate is similar to the one of flagless circuits.

\textbf{Definition - post-selected failure rate (PSFR):} The post-selected FR of a flag circuit is given by the number of measurements with undetected errors divided by the number of measurements where the flags indicated no error was present. In other words, missed failures divided by the total number of missed failures and the number of trials in which no failures existed.

\subsection{Adding Flags}
\label{sec:add}

One can formulate a variety of methods for adding flags to the circuit, with approaches ranging from exact methods (e.g. SAT-, SMT - solvers for solving the CSP problem formulation) to simple heuristics informed by straightforward cost metrics. 

\textbf{Problem Statement:} How can we evaluate the effectiveness of a flag scheme?

We focus on a heuristic for scalability purposes, while being aware that a more complex approach could lead to even lower failure rates. Our heuristic uses \emph{unique}, \emph{high weight} flags. In contrast to~\cite{debroy2020extended}, we use standard flags.

\textbf{Definition - unique flag:} If $A\{a_1, a_2...\}$ is the set of data qubits and $B\{b_1, b_2...\}$ is the set of flag qubits, let $C(a_i) \subset B$ be the set of flag qubits which touch $a_i$. With unique flags it follows that
\[
    |C(a_i)| \leq 1, \forall a_i \in A
\]
Meaning each data qubit is touching at most one flag qubit. Unique flags were used in order to avoid two or more independent flag errors being copied to the same data qubit.

\textbf{Definition - flag weight:} is practically the weight of the failure against which the flag is protecting. This is equal to the number of CNOTs surrounded by the flag.

When considering unique flags, their number scales in the number of qubits and not in the number of gates, because now there can be only one flag per qubit. For deep circuits, under this definition of uniqueness, temporal failure patterns will not be captured efficiently. Nevertheless, by replacing the ranking criteria to something similar to a \emph{failure hotspot}, one can add flags in the critical regions of the circuit~\cite{debroy2020extended}.

Equipped with the definition of weight and uniqueness, we compute a full list of unique candidate X-(Fig.~\ref{fig:exa}) and Z-flags (Fig.~\ref{fig:exz}) which protect CNOT sequences from the ICM circuit. Assuming the circuit has $g$ CNOTs, the number of candidate flags scales like $\mathcal{O}(g)$. We rank the candidate list based on the \emph{flag weight} and compile the flagged ICM circuit by including the best ranked flags. The resulting circuits will include one of the following: a) for resource efficiency reasons - a logarithmic number of flags $\mathcal{O}(\log g)$; b) for benchmarking purposes - a linear number of flags $\mathcal{O}(g)$.

\subsection{Examples}

We present examples of the previous definitions using Figure~\ref{fig:example}. The red X error, propagating through the CNOTs gates, is caught by the flag, and at the end of the circuit the error is present on the first, second and last qubit. Table~\ref{table:stabilizers} contains an exhaustive list of stabilizers computed for the error-free circuit, and for the circuit where a single X error is propagated to the outputs.

\textbf{Example - Failure:} A failure is a change in the outcome of stabilizer measurements between the fault-less and faulty circuit (marked red in Table~\ref{table:stabilizers}). For example, for the input state $\ket{000}$ the stabilizers are $\{+ZII, +IZI, +IIZ\}$ for the faultless circuit, and $\{-ZII, -IZI, +IIZ\}$ for the faulty. The error presented in the faulty circuit in Fig.~\ref{fig:exb} is an example of an error which is always caught by the flag. This is because in this case the sign of the flag stabilizers ($\pm IIIZ$) differs between the faultless and faulty circuit for all input states. However, for inputs $\ket{+0+}$ and $\ket{+++}$ this error is \emph{masked}: the rest of the stabilizers of the circuit are the same with and without the error. As a result the flags catch a "false positive" with these input states.

\textbf{Example - Failure Rate:} Continuing with Fig.~\ref{fig:exb}, let us consider running the circuit 800 times, such that for each of the 8 different input states there are 100 executions. Assuming that the error present in~Fig.\ref{fig:exb} occurs with probability $p=0.1$, then, on average, for each input the error occurs 10 times. There will be approximately 80 total cases with an error present. However, for $\ket{+0+}$ and $\ket{+++}$ the error is masked, such that there are in total only 60 cases with a failure. Thus, the failure rate of the flag-less circuit is $60/800=0.075$.

\textbf{Example - Post-Selected Failure Rate (PSFR):} What happens if a flag is added? In this case the flag from Fig.~\ref{fig:example} catches the illustrated error each time since the signs of the flag stabilizers with and without the error are different for all input states. This is true even in the case where the error is "masked" from the flagless stabilizers. This means that there are now 60 true positives (flag measurement correctly indicates an error is present), 20 false positives (flag measurement incorrectly indicates an error is present), 720 true negatives (flag measurement correctly indicates that no error is present) and 0 false negatives (flag measurement incorrectly indicates that no error is present) (Table~\ref{table:confusion}). With post-selection we discard all 80 cases where the flag indicates an error, and we are left with 720 circuit executions. Thus, the PSFR of the circuit is $0/(0+720)=0$ -- all circuits after post-selection are failure-free. Naturally, in a realistic scenario there are also errors which the flags do not catch, as well as new errors introduced by the flags such that the false negative rate and the PSFR would be nonzero.

\begin{table}[t!]
\centering
\begin{tabular}{| c | c | c |}
\cline{2-3}
\multicolumn{1}{c|}{} & \textbf{Failure present} & \textbf{No failure present} \\
\hline
\textbf{Flags on}  & True Positive (TP)  & False Positive (FP) \\
\hline
\textbf{Flags off} & False Negative (FN) & True Negative (TN) \\
\hline
\end{tabular}
\vspace{12pt}
\caption{Confusion matrix. The FR is defined as $(TP + FN) / (TP + FP + FN + TN)$. The PSFR is the FR where none of the "flags on" cases are considered, i.e. $ FN / (FN + TN)$.}
\label{table:confusion}
\end{table}

\subsection{Flag Fault-Tolerance: Upper Bound and Tuning}
\label{sec:tune}

Flag fine-tuning is based on the assumption that the flags are protected by e.g. a surface code and have a lower error rate compared to the data qubits.

\textbf{Definition:} The \emph{error multiplier} \textit{m} is the multiplicative factor that the base noise channel strength (i.e. error rate) $p_{ncs}$ of the data qubits is multiplied by to reach the error rate of the flag qubits. We assume it to be some non-negative number smaller than $1$. 

\textbf{Definition:} The \emph{circuit's failure rate upper bound} is achieved for perfect flags, meaning for $m=0$. In contrast, the \emph{lower bound} is achieved for $m=1$. For example, if $m=0.1$ then the flags would have $10$ times less failures compared to the data qubits of the circuit. 

The failure rate of the flag qubits is defined as  $ p_f = m \times p_{ncs}$. For example, if the depolarizing channel applied to data qubit gates has a strength of $p_{ncs}=10^{-4}$, and the failure rate of the flag qubits is reduced ten times, then the noise probability of the channel applied to the gates touching the flags is $p_f = 10^{-5}$.

\textbf{Problem Statement:} Assuming a given number of flags, find a value for the multiplicative factor \textit{m} such that the resulting circuit reaches some target failure rate $FR_{target}$. In the best case, $FR_{target}$ would be the upper bound.

We can frame this as a constraint satisfaction problem (CSP) having the constraints as the target failure rate $FR_{\text{target}}$ and the maximum number of flags $f_{max}$. The variables are the number of flags $f$ and the value of the error multiplier $m$. Instead of a CSP solution, we propose a heuristic (Alg.~\ref{alg:csp}) and construct a binary-search based algorithm. The complexity of our heuristic is $\mathcal{O}(f_{max}\log M)$, because we are adding at most $f_{max}$ flags, and we are selecting $m$ from the $M$ leafs of a binary tree obtained by splitting the interval $(0,1)$. Our algorithm is not guaranteed to find a solution even if we would allow arbitrary values of $M$, but empirical evaluations have shown that it converges very fast in practice.

\begin{algorithm}[!t]
\KwIn{$f_{max}$,  $FR_{target}$, $C$}
$f \gets 1$\;
$a \gets 0; b \gets 1; m \gets 0.5$\;

\While{$|FR - FR_{target}| > \epsilon$ and $f \leq min(f_{max}, n)$}{
    
    $L_u \gets$ All ways for adding flags to $C$;

    $L_c \gets $ Sort $L_u$ in descending order based on the number of gates between the flag CNOTs;

    Compile $C_f$ using the $f$ flags from $L_c$;
       
    \While{$|FR - FR_{target}| > \epsilon$}{
        $m \gets (a+b)/2$\;
        
        $FR \gets $ Simulate $C_f$ for current $m$\;
        
        \eIf{$FR - FR_{target} > \epsilon$}{
            $a \gets m$\;
        }{
            $b \gets m$\;
        }
    }
    $f \gets f + 1$\;
}
\caption{The flag failure rate multiplier $m$ and the number of flags $f$ needed is determined for the constraints $f_{max}$ (maximum allowed number of flags) and $FR_{target}$ (target failure rate). The algorithm is operating on a circuit $C$ of $N$ qubits, and the current failure rate $FR$ has to be within $\epsilon$ from the target failure rate. The value of $m$ is chosen from an interval $(a,b)$ and making sure that $a$ and $b$ are lower bounded by a threshold value that limits the depth of binary search.}
\label{alg:csp}
\end{algorithm}

\begin{figure*}[!t]
    \begin{subfigure}{0.32\linewidth}
        \centering
        \includegraphics[width=\columnwidth]{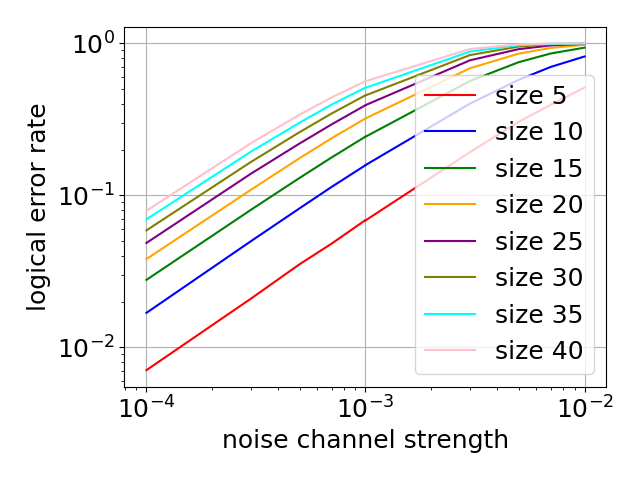}
        \caption{}
        \label{fig:basic}
    \end{subfigure}
     \begin{subfigure}{0.32\linewidth}
        \centering
        \includegraphics[width=\columnwidth]{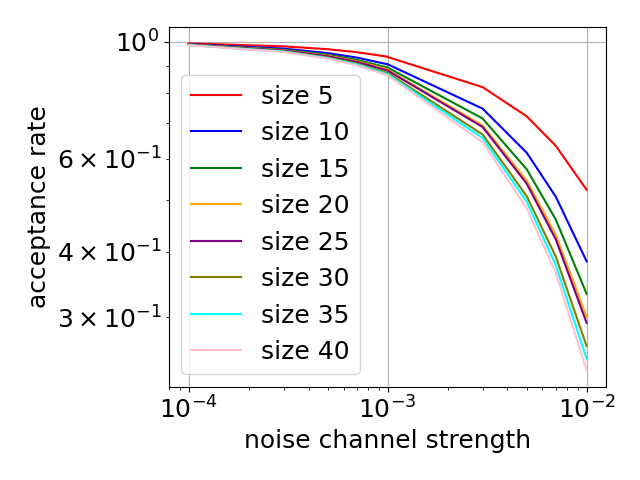}
        \caption{}
        \label{fig:acceptance}
    \end{subfigure}
    \begin{subfigure}{0.32\linewidth}
        \centering
        \includegraphics[width=\columnwidth]{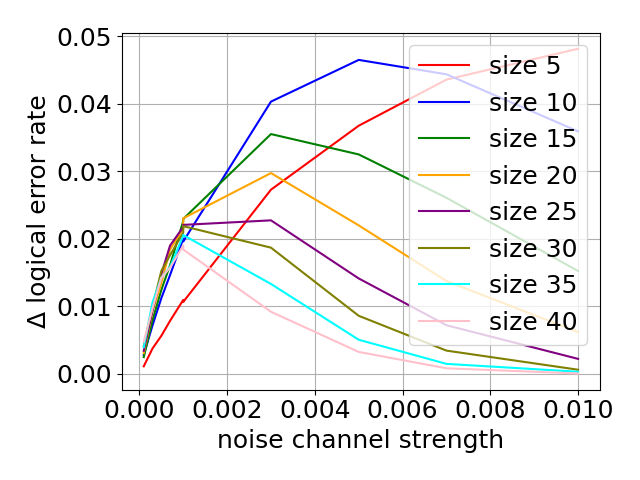}
        \caption{}
        \label{fig:error_delta}
    \end{subfigure}

    \begin{subfigure}{0.32\linewidth}
        \centering
        \includegraphics[width=\columnwidth]{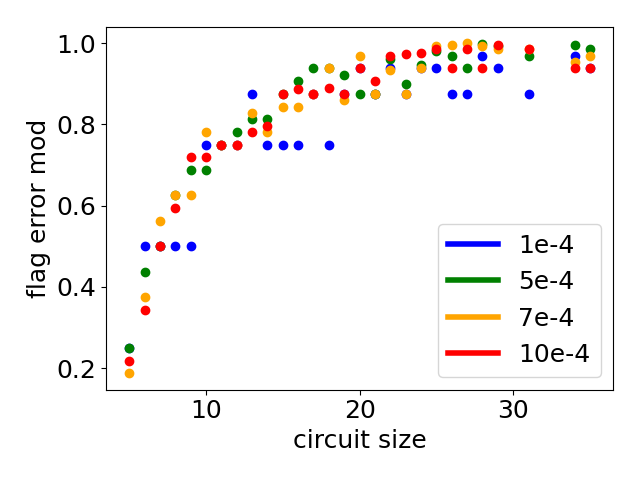}
        \caption{}
        \label{fig:error_mod_vs_size}
    \end{subfigure}
    \begin{subfigure}{0.32\linewidth}
        \centering
        \includegraphics[width=\columnwidth]{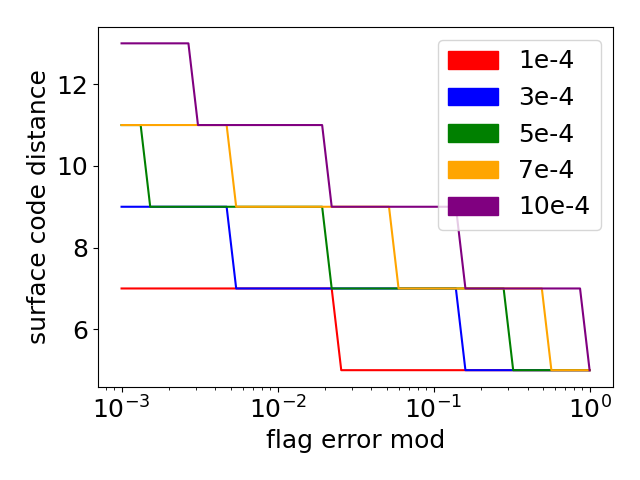}
        \caption{}
        \label{fig:5log_surface_d}
    \end{subfigure}
    \begin{subfigure}{0.32\linewidth}
        \centering
        \includegraphics[width=\columnwidth]{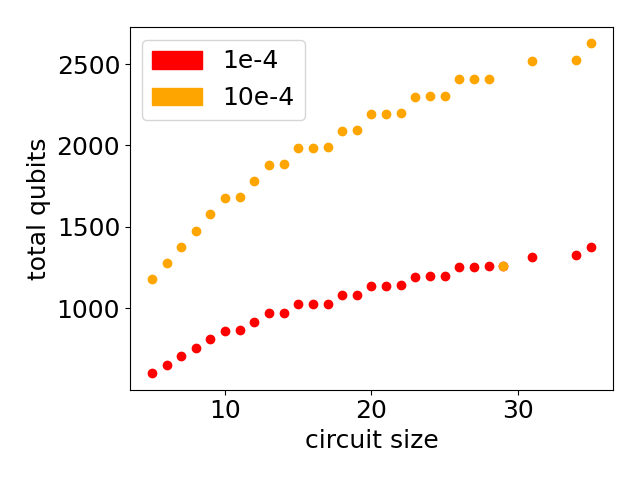}
        \caption{}
        \label{fig:5log_surface_q}
    \end{subfigure}
    
    \caption{Benchmarking results for using $5 \log_{2}(N)$ flags, where \textit{N} is the size of the adder-like circuit. The horizontal axis is the noise channel strength i.e. the error rate at each gate. 
    (a) The failure rate after post selection the flags; (b) The acceptance rate i.e. what percentage of runs is accepted during post selection using the flags; (c) The difference in failure rate, i.e. the difference between the failure rate of the flagless ICM adder and the flag adder of the same size.
(d) Error multiplier \textit{m} needed to reach $N-1$ target failure rate for different base error rates $p_{ncs}$ (colors). (e) Surface code distance needed to achieve the error rate of the horizontal axis for different $p_{ncs}$ (f) Number of qubits required (incl. surface code) to achieve the failure rate of a size $N-1$ circuit using a circuit of size \textit{N} considering the $p_{ncs}$ from the colored lines. e.g. circuit size $N=35$ needs around 2500 qubits to get to the FR of $N-1=34$ when considering that the $p_{ncs}$ is 0.001 (yellow).}
\end{figure*}

\section{Results and Analysis}
\label{sec:res}

First, we examine the lower and upper bound failure rates, as the effect of changing the value of \textit{m} (Sec.~\ref{sec:improve}). Second, we evaluate the inner while-loop of Alg.~\ref{alg:csp}: after specifying $FR_{target}$ based on the circuit sizes, we can find values of \textit{m} for each circuit (Sec.~\ref{sec:m}). Finally, we estimate the physical resources required based on these results, and look at how many physical qubits are needed in order to provide sufficiently low error rates on the flags using surface codes (Sec.~\ref{sec:sc}).

\subsection{Benchmark Environment}
\label{sec:circuits}

All benchmark circuits were in ICM form~\cite{vijayan2024compilation}(Fig.~\ref{fig:ICM}). We focused on adders of different sizes as they share a similar structure and can be better compared to each other. Significant changes to the adder circuits were made and as a result the final circuits used were no longer adders but rather adder-like circuits. These still retain the characteristic of having a similar structure regardless of size.

We start from adder circuits obtained from Qualtran~\cite{harrigan2024expressing}. These circuits are modified, resulting in CNOT-only circuits with a structure similar to the original adders. The resulting circuits were not optimized. We obtained the adder-like circuits using the following procedure: 1) compile to Clifford+T; 2) remove T-gates and Hadamard gates; 3) generate the ICM form of the corresponding Clifford; 4) remove classical controls and readouts. We decided to completely remove T and H gates, instead of replacing them with an ancilla for gate teleportation, as this would have led to a significantly increased qubit count.

Numerical estimation of the circuits' FR was performed by simulating the circuit for $10^4$ times using Stim~\cite{gidney2021stim} for 100 different input states as discussed below, such that the total number of simulation runs for any given circuit was $10^6$.

The number of input states for a circuit was either the number of total possible input states or 100, whichever is smaller. The inputs were randomly determined by generating a maximum of 100 random input strings for each circuit. Based on the input string each qubit was prepared in either $\ket{0}$, or $\ket{+}$ state by adding a Hadamard gate. Similar approaches for using randomly selected inputs as stimuli for the circuits were used in the context of quantum circuit equivalence checking~\cite{burgholzer2021random}.

We assume the qubit initialization and measurements, including flags, at the end of the circuit to be noiseless. The two qubit gates (CNOTs) have a depolarizing probability of failure $p_{ncs}$.

\subsection{Improvement from Flagless Circuits}
\label{sec:improve}

How much better are the flagged circuits in comparison to the original flagless ICM circuits if we keep some, or all of the unique flags suggested by the heuristic?
\begin{figure}[!h]
    \centering
    \includegraphics[width=0.7\columnwidth]{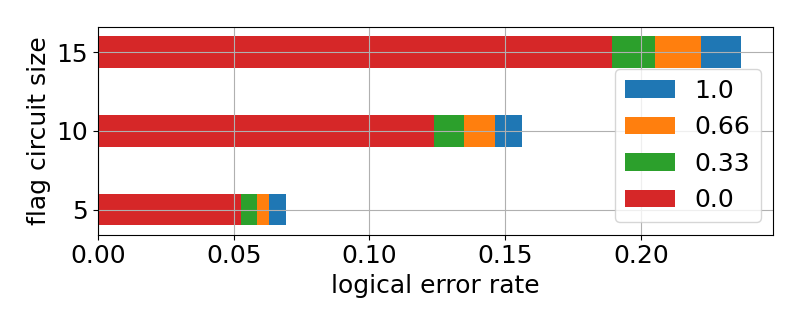}
    \caption{Improving a circuit's total failure rate (horizontal axis) by lowering only the flags' failure rate. The error multiplier (color coded) is the factor by which the error rate of the gates attached to the flag qubits is multiplied by. The vertical axis is the size of the adder. Results for perfect flags are red.}
    \label{fig:error_mod}
\end{figure}

\textbf{Result:} Fault-tolerant flags can be used to improve a circuit's failure rate. If \textit{m=0}, we can consider the case that the flags are perfect i.e. they do not introduce any new errors into the circuit. This gives us a best case scenario for how much the failure rate of the circuit can be improved by adding flags according to the heuristic used (Fig.~\ref{fig:error_mod}). 

This result does not only depend on the amount of flags but it is also dependent on both the size of the adder as well as the noise channel strength, i.e. physical error rate.

\textbf{Result:} For each circuit size, there is a regime of noise channel strength where the benefit of using flags is at a maximum. The amount of flags used does not change the location of this peak, but rather affects its size (Fig.~\ref{fig:error_delta}). The location of the peak moves towards smaller values of the noise channel strength as the size of the circuit increases.

\subsection{Fault-Tolerant Flags}
\label{sec:m}

Based on the previous observations, lowering \textit{m}, i.e. increasing the error correction on the flags, also lowers the failure rate of the entire circuit. However, what would be an optimal value of $m$? In the case of adder-like circuits, we wish to tune the value of \textit{m} such that an \textit{N}-circuit reaches a target  $FR_{target}$ equivalent to the failure rate of a $\textit{N} - 1$ circuit. For example, for a circuit of size 10 the target failure rate with flags would be the flagless failure rate obtained with a circuit of size 9. This approach is motivated by the fact that quite many quantum circuits have a gate count which scales according to some function that has the number of qubits as a parameter (e.g. the QFT and adders~\cite{gidney2021factor}, QRAM~\cite{jaques2023qram}, error-correction circuits~\cite{lidar2013quantum}).

Let us examine the case where we have $5\log_{2}{N}$ flags for each circuit size $N$. In Fig.~\ref{fig:error_mod_vs_size} we have plotted the error multiplier values needed to reach $N-1$ target failure rate for different values of $p_{ncs}$ and circuit sizes. 
This was done using Alg.~\ref{alg:csp}: we searched for a value for \textit{m} between 0 and 1, and the difference allowed from the target error rate was $\epsilon = 0.0005$.

\textbf{Result:} From the results in Fig.~\ref{fig:error_mod_vs_size} we can see that regardless of the base error $p_{ncs}$ the error multiplier \textit{m} vs circuit size \textit{N} curve follows roughly the same shape: with small circuits \textit{m} must be small in order to reach $N-1$ target failure rate. For large circuits, \textit{m} does not have to be as small to achieve the same thing. This means that for smaller circuits the flags must have more error correction than if the circuits were larger.

\subsection{Surface Code Flags}
\label{sec:sc} 

If we assume that the flags have an improved error rate due to the fact that they are protected by a surface code, what would the distance of the surface codes be in order to achieve these error multiplier values? Based on~\cite{o2025compare} we calculated the required distances using $p_f = 0.08*(p_{ncs}/0.0053)^{0.58d - 0.27}$ where $d$ is the distance of the surface code. We assumed that the surface codes are all rotated, such that per logical qubit there are $2d^2 - 1$ physical qubits which include data and syndrome ones. The resulting number of qubits needed are illustrated in Fig.~\ref{fig:5log_surface_d}).

Based on the distance required for the surface codes protecting the flags, we can estimate on how many qubits would be required in total to build such circuits. It is worth noting that these estimations do not include the qubits needed to error correct the two qubit gates (i.e. CNOTs) between the flags and the data qubits. Therefore, the estimations presented here are a lower bound. The results of these calculations are plotted in Fig.~\ref{fig:5log_surface_q}.

\section{Conclusion}
\label{sec:concl}

We presented \textsc{Medusa}, a scalable method for lowering the failure rate of large ICM circuits by adding unique flags and tuning their fault-tolerance. Our simulations show that even moderate improvements in the reliability of flag qubits can reduce the overall circuit FR, and that for adder-like circuits a small number of well-placed flags can make a large circuit behave like a smaller one in terms of logical error rate. 

Similarly to~\cite{debroy2020extended}, we conclude that the position of the flags is important in increasing the fault-tolerance, and improved insertion schemes are needed.
Additionally, our method assumes ICM circuits. Due to their CNOT-only structure, by carefully analyzing the logical correlations existing within the circuits~\cite{zhou2024algorithmic}, we conjecture that larger improvements of fault-tolerance are possible. 

By relating the required flag reliability to surface-code distances, we quantify the physical resources needed to achieve a target failure rate. These results suggest that partially protected flag qubits offer a  cost-effective alternative to fully fault-tolerant syndrome extraction. Future work includes refining flag-placement heuristics and extending the method to multi-flag gadgets.

\section*{Acknowledgements}
The calculations presented above were performed using computer resources within the Aalto University School of Science “Science-IT” project. This research was developed in part with funding from the Defense Advanced Research Projects Agency [under the Quantum Benchmarking (QB) program under award no. HR00112230007 and HR001121S0026 contracts], and was supported by the QuantERA grant EQUIP through the Academy of Finland, decision number 352188. The views, opinions and/or findings expressed are those of the author(s) and should not be interpreted as representing the official views or policies of the Department of Defense or the U.S. Government.

\balance

\bibliographystyle{acm}
\bibliography{__main}

\end{document}